\documentclass[times,twocolumn,final]{elsarticle}

\usepackage{amssymb}

\usepackage{natbib}
\usepackage{hyperref}

\usepackage{amsmath,amssymb,amsfonts}
\usepackage{algorithmic}
\usepackage{graphicx}
\usepackage{textcomp}

\usepackage[ruled]{algorithm2e}
\SetKwInput{KwRequire}{Require}

\usepackage{makecell}
\usepackage{bm}
\usepackage{array}
\usepackage{threeparttable}
\usepackage{color}
\usepackage{xcolor}
\definecolor{R}{rgb}{1,0,0}
\definecolor{B}{HTML}{000000}
\definecolor{G}{HTML}{000000}
\definecolor{O}{HTML}{000000}
\definecolor{P}{HTML}{000000}

\usepackage{booktabs,diagbox,float}

\journal{Expert Systems With Applications}

\begin{document}

\begin{frontmatter}



\title{Contextual Embedding Learning to Enhance 2D Networks for\\Volumetric Image Segmentation}


\author[1,2,3]{Zhuoyuan Wang}
\author[4]{Dong Sun}
\author[1,2,3]{Xiangyun Zeng}
\author[5]{Ruodai Wu\corref{cor1}}
\author[1,2,3]{Yi Wang\corref{cor1}}
\cortext[cor1]{Corresponding authors: Ruodai Wu and Yi Wang\\
\textit{Email addresses}:
2018222020@email.szu.edu.cn (Zhuoyuan Wang),
sundong22@huawei.com (Dong Sun),
1910242017@email.szu.edu.cn (Xiangyun Zeng),
wuruodai@szu.edu.cn (Ruodai Wu),
onewang@szu.edu.cn (Yi Wang).}
\address[1]{National-Regional Key Technology Engineering Laboratory for Medical Ultrasound, Guangdong Key Laboratory for Biomedical Measurements and Ultrasound Imaging, School of Biomedical Engineering, Shenzhen University Medical School, Shenzhen University, Shenzhen, China}
\address[2]{Smart Medical Imaging, Learning and Engineering (SMILE) Lab, Shenzhen, China}
\address[3]{Medical UltraSound Image Computing (MUSIC) Lab, Shenzhen, China}
\address[4]{Huawei Cloud Computing Technologies Company Limited, China}
\address[5]{Department of Radiology, Shenzhen University General Hospital, Shenzhen University, Shenzhen, China}


\begin{abstract}
The segmentation of organs in volumetric medical images plays an important role in computer-aided diagnosis and treatment/surgery planning.
Conventional 2D convolutional neural networks (CNNs) can hardly exploit the spatial correlation of volumetric data.
Current 3D CNNs have the advantage to extract more powerful volumetric representations but they usually suffer from occupying excessive memory and computation nevertheless.
In this study we aim to enhance the 2D networks with contextual information for better volumetric image segmentation.
Accordingly, we propose a contextual embedding learning approach to facilitate 2D CNNs capturing spatial information properly.
Our approach leverages the learned embedding and the slice-wisely neighboring matching as a soft cue to guide the network.
In such a way, the contextual information can be transfered slice-by-slice thus boosting the volumetric representation of the network.
Experiments on challenging prostate MRI dataset (PROMISE12) and abdominal CT dataset (CHAOS) show that our contextual embedding learning can effectively leverage the inter-slice context and improve segmentation performance.
The proposed approach is a plug-and-play, and memory-efficient solution to enhance the 2D networks for volumetric segmentation.
\textcolor{G}{Our code is publicly available at~\url{https://github.com/JuliusWang-7/CE\_Block}.}
\end{abstract}




\end{frontmatter}


\section{Introduction}
\label{sec:introduction}
The segmentation of organs in volumetric medical images (e.g., computed tomography (CT), magnetic resonance imaging (MRI), 3D ultrasound) enables quantitative analysis of anatomical parameters such as shape, boundary, volume, etc~\citep{shen2017deep, litjens2017survey, zhang2024deep}.
Also, the segmentation is often the very first step in the workflow of other medical image computing tasks, such as computer-assisted diagnosis~\citep{hesse2020intensity, wang2019deeply, orlando2022effect}, and surface-based registration~\citep{wang2024recursive, ghavami2019automatic, wang2017online}.
However, it is tedious and time-consuming for the clinicians to manually delineate the organ's contour.
Therefore, the automatic segmentation approaches are highly needed to satisfy the constantly increasing clinical demand.

\begin{figure*}[t]
	\centering
	\includegraphics[width=0.99\linewidth]{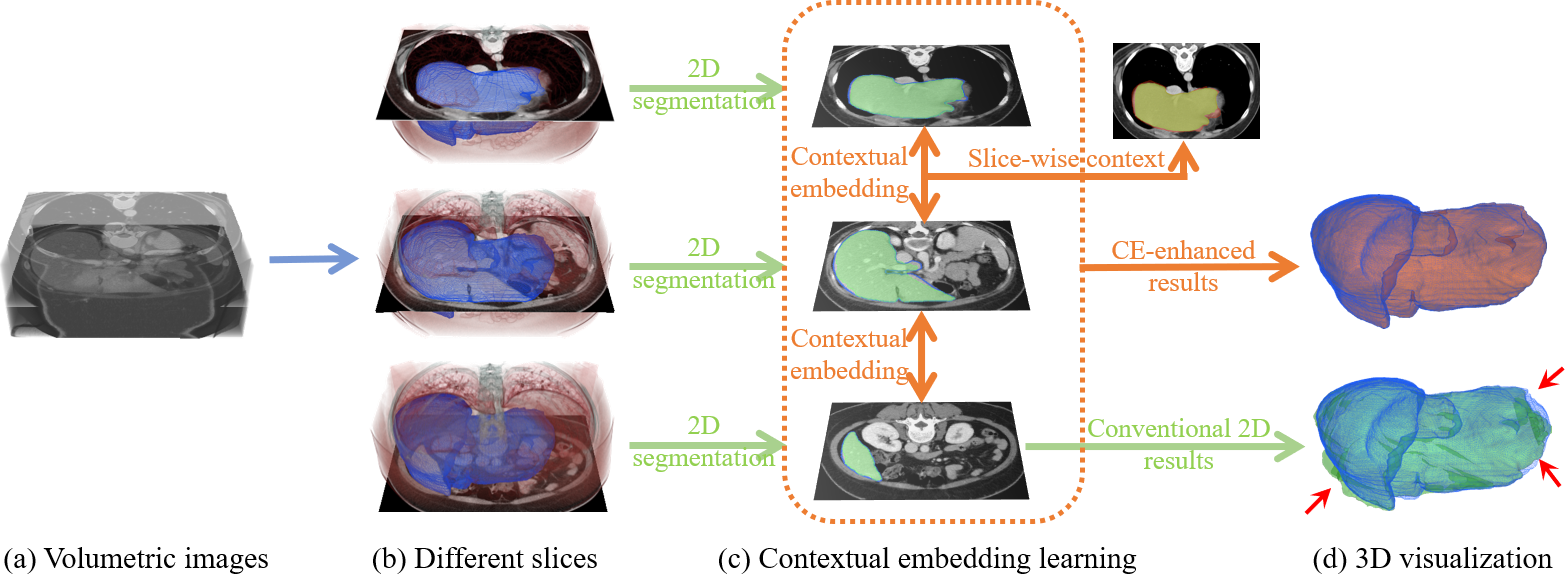}
	\caption{For volumetric medical images, conventional 2D networks can only segment each 2D slice individually, but can hardly obtain the context information between slices, which results in incomplete and discontinuous segmentation results as shown in (d).
	Our contextual embedding (CE) approach transfers contextual information via a slice-wisely neighboring matching mechanism thus boosting the volumetric representation of the 2D network.
	In (d), the blue, green, and orange surfaces indicate ground-truth segmentation, conventional 2D result and CE-enhanced result, respectively.
	It shows our approach makes the segmentation more smoother and complete in 3D.
	}
	\label{fig:intro}
\end{figure*}

A large number of literatures have been presented the widest variety of methodologies in medical image segmentation.
Early approaches mainly rely on thresholding, edge detection, atlas matching, deformable models, and machine learning techniques~\citep{pham2000current}.
Although these approaches have shown their successes in certain circumstances, medical image segmentation is still challenging mainly due to the difficulty of designing and extracting discriminative features.

With the promising capability of deep learning techniques in the society of computer vision and pattern recognition~\citep{lecun2015deep}, convolutional neural networks (CNNs) have also become a primary option for medical image segmentation.
Various CNN architectures have achieved excellent segmentation results in many medical applications.
One of the most well-known CNN architectures for medical image segmentation is the U-net~\citep{ronneberger2015u}, which adopts the skip connections to effectively aggregate low-level and high-level features.
The U-net has now been the benchmark for lots of segmentation tasks and based on which many 2D variants have been devised to target specific applications.
Chen~\textit{et al.}~\citep{Chen_2016_CVPR} proposed a multi-task learning network for gland segmentation from histopathological images, which explores multi-level contextual features from the auxiliary classifier on top of the U-net structure.
Ibtehaz~\textit{et al.}~\citep{ibtehaz2020multiresunet} took inspiration from Inception blocks~\citep{szegedy2017inception} and proposed MultiRes blocks to enhance U-Net with the ability of multi-resolutional analysis.
Zhou~\textit{et al.}~\citep{zhou2018unet++} re-designed the skip connections and proposed a UNet++ which is a deeply-supervised encoder-decoder architecture with nested and dense skip connections.
Jha~\textit{et al.}~\citep{8959021} proposed ResUNet++, which took advantage of residual blocks, squeeze and excitation blocks, and strous spatial pyramidal pooling (ASPP) to segment the colorectal polyps.
\textcolor{O}{Gao~\textit{et al.}~\citep{gao2021utnet} introduced UTNet, a hybrid Transformer architecture that blended self-attention mechanism with CNNs to enhance medical image segmentation.}
However, directly employing 2D networks to deal with 3D images in a slice-by-slice way may result in discontinuous or inaccurate segmentation in 3D space (see the conventional 2D results in Fig.~\ref{fig:intro}), due to the limited volumetric representation capability of 2D learning~\citep{8379359}.
To better extract spatial information, some 2.5D approaches have been proposed~\citep{prasoon2013deep, wang2017multi, vesal2022domain}.
In general, these approaches still employ 2D convolutional kernels, whereas extracting features from three orthogonal views, i.e., transversal, coronal, and sagittal planes.
Although 2.5D approaches could slightly improve the 2D networks' results, they still may not well exploit the original volumetric images~\citep{kamnitsas2017efficient}.
Moreover, the three branches of 2.5D networks would cause a rapid increase in parameters and memory usage.

To better empower the networks with richer volumetric information, {\c{C}}i{\c{c}}ek~\textit{et al.}~\citep{cciccek20163d} proposed 3D U-net to extend U-net architecture directly processing 3D images.
However, the high memory consumption of 3D U-net would limit the depth of networks.
Milletari~\textit{et al.}~\citep{milletari2016v} proposed another 3D derivation of U-net architecture, named V-net, applying residual connections to enable a deeper network.
Due to the advantages of residual connections, several studies~\citep{chen2018voxresnet, wang20203d, wang2021rar} further employed the residual design to boost the 3D networks.
Furthermore, to better aggregate the multi-scale spatial features, various attention mechanisms have been added into the 3D CNNs~\citep{oktay2018attention, wang2019deep, lin2021variance, huang2023joint, ates2023dual, yang2024recurrent}. \textcolor{O}{Hatamizadeh~\textit{et al.}~\citep{hatamizadeh2022unetr} presented a novel architecture, called UNETR (UNET Transformers), which employed a Transformer as the encoder to acquire sequence representations of the input volume and captured global multi-scale information.}
Although 3D CNNs are beneficial to extract more powerful volumetric features, they still suffer from occupying excessive memory and computation.

It is worth noting that combining CNNs with Recurrent Neural Networks (RNNs), such as Long Short-Term Memory (LSTM)~\citep{hochreiter1997long} and convolutional-LSTM~\citep{xingjian2015convolutional}, can obtain slice-wise contextual information.
Chen~\textit{et al.}~\citep{chen2016combining} combined a 2D U-net with bi-directional LSTM-RNNs to mine the intra-slice and inter-slice contexts for 3D image segmentation.
Poudel~\textit{et al.}~\citep{poudel2016recurrent} devised a recurrent fully-convolutional network to learn inter-slice spatial dependences for cardiac MRI segmentation.
Alom~\textit{et al.}~\citep{Alom2018Recurrent} developed a recurrent residual U-net model (R2U-Net), which utilized the recurrent residual convolutional layers to ensure better feature representation for segmentation tasks.
Novikov~\textit{et al.}~\citep{novikov2018deep} proposed Sensor3D to integrate bi-directional C-LSTMs into a U-net-like architecture to extract the inter-slice context sequentially, which achieved notable performance in CT segmentation.
\textcolor{O}{Kang~\textit{et al.}~\citep{kang2022renal} combined 3D CNN with ConvLSTM for fine segmentation, aiming to better capture contextual information between slices and thereby improving segmentation results.}
RNNs can extract the contextual relationship, yet they bring extra computational cost and memory usage.

In this study, we aim to enhance the 2D networks with contextual information for the accurate and efficient volumetric image segmentation.
Our proposed network shares the backbone features with the encoder of conventional 2D segmentation networks and utilizes the slice-wise spatial context adequately by the proposed contextual embedding learning approach, thus improving segmentation performance (see Fig.~\ref{fig:intro}).
The contributions of our network are as follows:
\begin{itemize}
\item We provide a contextual embedding learning approach, which leverages the learned embedding and the slice-wisely neighboring matching as a soft cue to guide the network.
In such a way, the contextual information can be transferred slice-by-slice thus facilitating 2D networks to capture volumetric representation properly.
\item The proposed contextual embedding block is a plug-and-play, and memory-efficient solution to enhance the 2D networks for volumetric segmentation.
\item Experiments on challenging prostate MRI dataset and abdominal CT dataset demonstrate that our approach consistently enhances 2D networks' performance on volumetric images.
In addition, our approach is more lightweight compared with the 3D networks.
\end{itemize}

The rest of this article is organized as follows.
Section~\ref{sec:method} introduces our segmentation network and further elaborates the proposed contextual embedding block.
Section~\ref{sec:exper} shows the segmentation performance of our network.
Sections~\ref{sec:discussion} and~\ref{sec:con} present the discussion and conclusion of this study, respectively.

\begin{figure*}[t]
	\centering
	\includegraphics[width=0.99\linewidth]{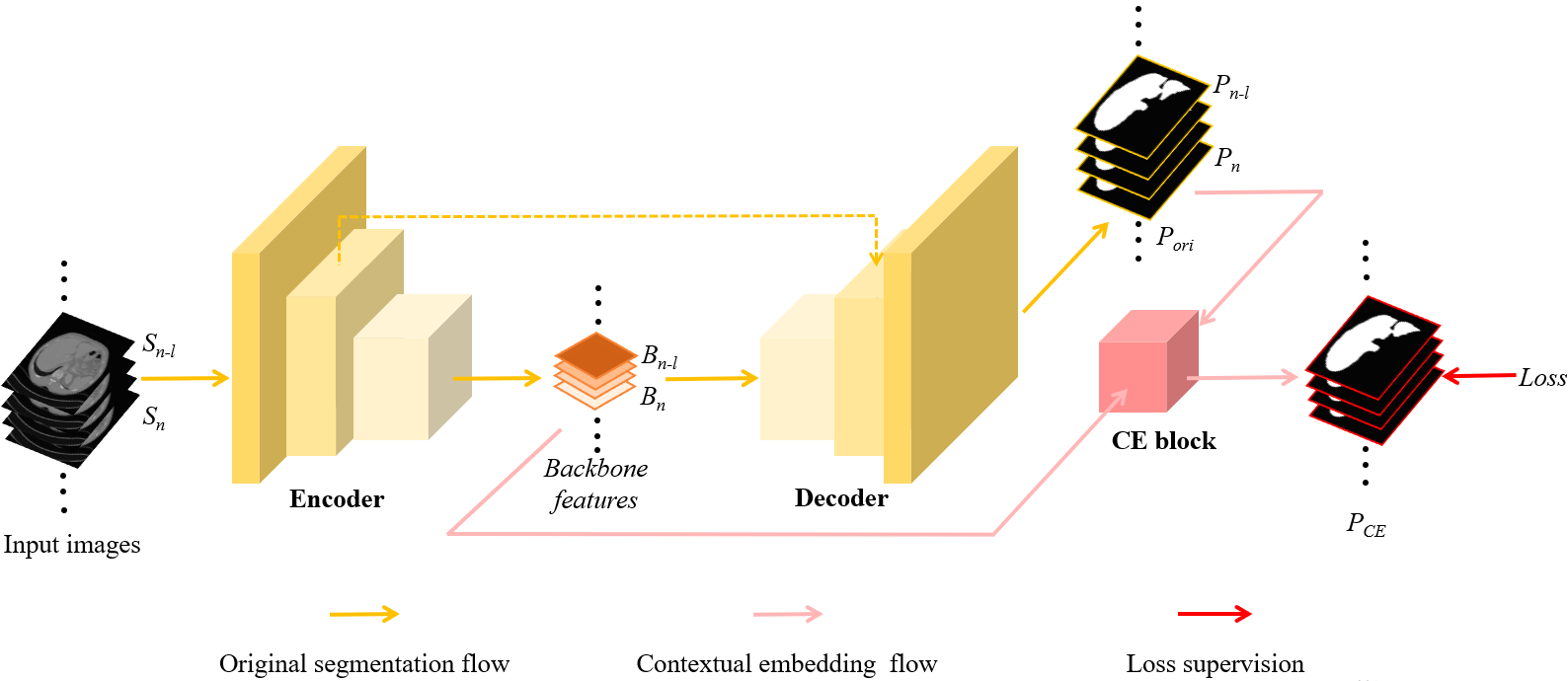}
	\caption{The schematic overview of the proposed network.
	The yellow arrows indicate the workflow of the conventional 2D segmentation using the encoder-decoder architecture, whereas the pink flow shows the plug-and-play contextual embedding (CE) block to enhance the volumetric representation of the 2D network.
	Specifically, the CE block leverages the prediction of the neighboring slice (i.e., the $P_{n-l}$ of the $S_{n-l}$) to calculate a distance map by matching the current slice embedding to the embedding of the neighboring slice (see details in Fig.~\ref{fig:distancemap}). 
	Then the CE block aggregates the neighboring matching distance map, the prediction of the neighboring slice ($P_{n-l}$), the original prediction of the current slice ($P_{n}$), and the backbone feature ($B_{n}$) to generate the refined prediction ($P_{n-CE}$) of the current slice ($S_{n}$).}
\label{fig:structure}
\end{figure*}

\section{Methods}
\label{sec:method}
Fig.~\ref{fig:structure} illustrates how the proposed contextual embedding (CE) block works in an encoder-decoder segmentation architecture.
The backbone features are shared by the decoder and the CE block.
In addition to the backbone features, the original prediction set ($P_{ori}$) from the decoder is the other input branch for the CE block.
The CE block utilizes the context between neighboring slices as soft cue to track the variation of the three-dimensional shape in the axial direction to generate the final prediction set $P_{CE}$.

\subsection{Overview of the Contextual Embedding Block}
\label{ssec:stm}  
Since conventional 2D CNNs can only perform convolutions in the axial plane, they can hardly obtain the slice-wise information that characterizes the implicit dependency between slices.
In such a case, 2D segmentation results may appear discontinuous or incomplete in 3D space.
In contrast, our designed CE block propagates the implicit dependency from neighboring slice to the current slice with a memory-efficient manner to improve the segmentation results.

\begin{figure*}[t]
	\centering
	\includegraphics[width=0.99\linewidth]{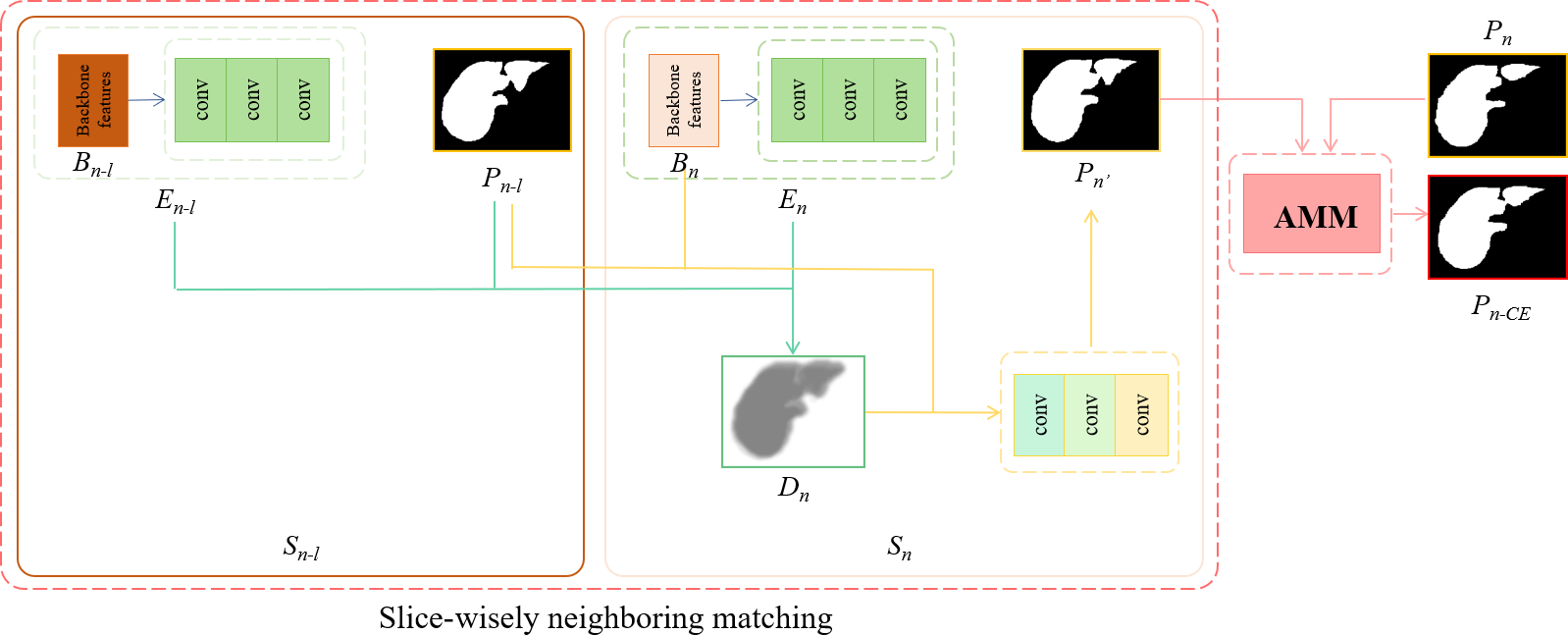}
	\caption{The detailed design of the contextual embedding block.
	In order to segment the current slice ($S_n$), the backbone features ($B$), embedding vectors ($E$), and the prediction of the neighboring slice ($P_{n-l}$ of $S_{n-l}$) are employed for it.
	First, a distance map ($D_{n}$) is obtained by matching the current slice embedding ($E_n$) to the embedding of the neighboring slice ($E_{n-l}$), see the green flow.
	Then, the $D_{n}$, together with the prediction of the neighboring slice ($P_{n-l}$) and the backbone feature ($B_{n}$) are combined to generate the new prediction of the current slice ($P_{n'}$), see the yellow flow.
	Finally, the $P_{n'}$ and the original prediction of the current slice ($P_{n}$) are aggregated through an attention merge module (AMM) to produce the final segmentation.
	The details of the AMM is shown in Fig.~\ref{fig:merge}.}
	\label{fig:distancemap}
\end{figure*}

Fig.~\ref{fig:distancemap} illustrates the detailed design of the proposed contextual embedding block.
Compared with the conventional networks only leveraging backbone features for the segmentation prediction,
we also employ the original predictions as the additional cue to generate the segmentation with \textcolor{O}{slice-wise} context.
To segment the current slice, the CE block leverages the prediction of the neighboring slice to calculate a distance map by matching the current slice embedding to the embedding of the neighboring slice (see details in Section~\ref{ssec:nm}).
Then the CE block combines the neighboring matching distance map, the prediction of the neighboring slice, and the backbone features to generate the refined prediction of the current slice.
Finally, the refined prediction and the original prediction of the same slice are aggregated through an attention merge module (AMM) to produce the final segmentation (see details in Section~\ref{ssec:amm}).
Note that the embedding and the neighboring matching mechanism are used as the soft cue to guide the network, thus the whole network can be trained end-to-end without an extra loss on the embedding.

\subsection{Embedding Space}
\label{ssec:el} 
For each pixel $p$ from a slice $S$, we obtain its corresponding embedding vector $e_{p}$ in the learned embedding space.
Specifically, as shown in Fig.~\ref{fig:distancemap}, three consecutive convolutional blocks are employed to non-linearly map backbone features ($B$) to the embedding space ($E$).
Each convolutional block is composed of a convolutional layer and a batch normalization layer followed by a rectified linear unit (ReLU) activation layer.

In the embedding space, the pixels from the same class will be close whereas pixels from different classes will be far way~\citep{chen2018blazingly}.
Thus we utilize the distance of pixels in the embedding space as a soft cue to guide the network.
As described in~\citep{fathi2017semantic}, we represent the distance between pixels $p$ and $q$ in the embedding space as
\begin{equation}
\label{formula:distanceP2P}
d(p, q)=1-\frac{2}{1 + \exp(\left \| e_{p}-e_{q} \right \|^{2})} ,
\end{equation}
where $e_{p}$ and $e_{q}$ are the corresponding embedding vectors of $p$ and $q$, respectively.
The embedding distance $d(p, q)$ ranges from 0 to 1, depending on the similarity between pixels $p$ and $q$.
For similar pixels, the $d(p, q)$ approaches to 0, and for pixels which are far away in the embedding space, the $d(p, q)$ will be close to 1.

\subsection{Slice-wisely Neighboring Matching}
\label{ssec:nm} 
To segment the current slice $S_{n}$, we transfer the context information from the neighboring slice $S_{n-l}$ to $S_{n}$ by considering the slice-wisely neighboring matching in the learned embedding space\footnote{$l$ denotes the interval between the current and neighboring slices. Note that for the first $l$ slices, we transfer the information from $S_{n+l}$ to $S_{n}$.}.
Specifically, as shown in Fig.~\ref{fig:distancemap}, a neighboring matching distance map $D_{n}$ is calculated by using the embedding of current slice ($E_{n}$), the embedding ($E_{n-l}$) and prediction ($P_{n-l}$) of neighboring slice:
\begin{equation}
\label{formula:dismap}
D_n(p)=\min_{q\in P_{n-l, o}}d(p, q)
\end{equation}
where $p$ and $q$ represent the pixels from $S_{n}$ and $S_{n-l}$, respectively.
$D_n(p)$ indicates the value of matching distance at $p$ position.
$P_{n-l, o}$ denotes the set of pixels which belong to the organ area according to the prediction $P_{n-l}$.
The embedding distance $d(p, q)$ can be computed using Eq~(\ref{formula:distanceP2P}).
By analyzing Eq~(\ref{formula:dismap}), it can be observed that the $D_n(p)$ affords for each pixel from the current slice a soft cue of how likely it belongs to the organ area.

\begin{figure}[t]
	\centering
	\includegraphics[width=0.99\linewidth]{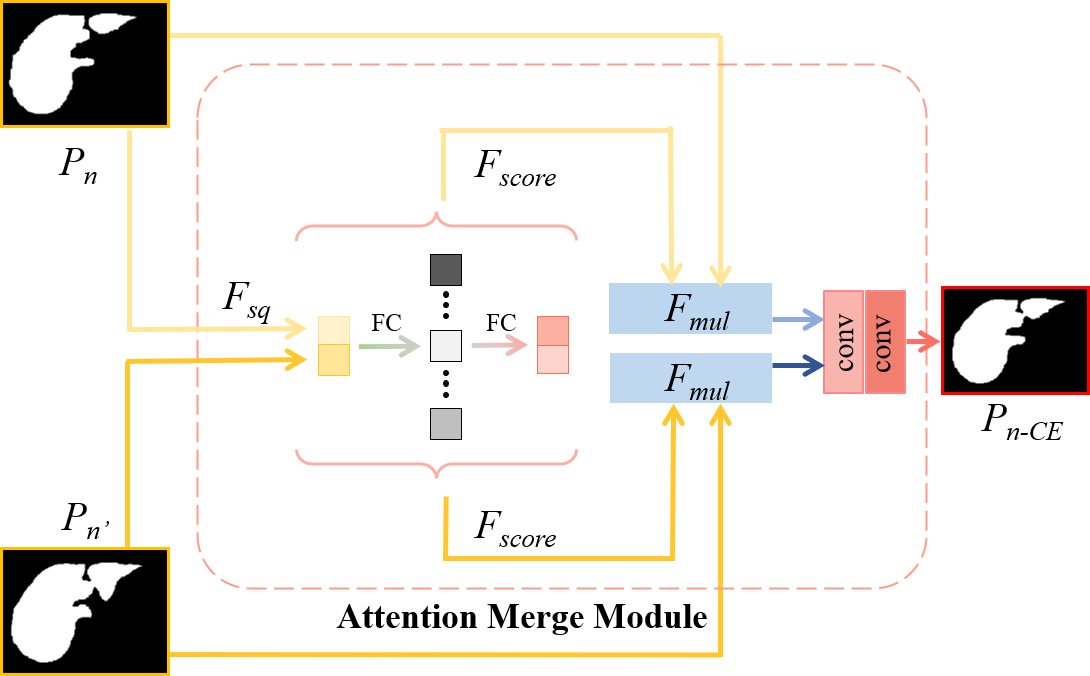}
	\caption{The structure of the attention merge module (AMM).
	The new prediction $P_{n'}$ and the original prediction $P_{n}$ of the current slice are aggregated through the AMM. 
	The AMM consists of three sequential functions of $F_{sq}$, $F_{score}$, $F_{mul}$, and the segmentation convolutions to fuse information and generate the final prediction $P_{n-CE}$.
	}
	\label{fig:merge}
\end{figure}

When computing Eq~(\ref{formula:dismap}), each pixel from the current slice has to be compared with the pixels from the neighboring slice, which may result in false positive matches and redundant computations~\citep{voigtlaender2019feelvos}.
Therefore, in practice we do not compute $D_n$ in a global manner but instead we apply a local matching as described in FlowNet~\citep{Dosovitskiy2015FlowNet}.
For pixel $p$ from the current slice, we only compare pixel $q$ from neighboring slice in a local patch with size $k$ (i.e., the set of pixels which are at most $k$ pixels away from pixel $p$ in both $x$ and $y$ direction).
In such a way, the distance map $D_n$ can be implemented using efficient and fast matrix operations without the brute force search algorithm.

We then concatenate the distance map $D_{n}$, together with the prediction of the neighboring slice ($P_{n-l}$) and the backbone feature ($B_{n}$), and feed them into the segmentation convolutions to generate the new prediction of the current slice ($P_{n'}$), see the yellow flow in Fig.~\ref{fig:distancemap}.
The $P_{n'}$ aggregates the context information of neighboring slice to predict the current slice.

\begin{figure}[t]
	\centering
	\includegraphics[width=0.99\linewidth]{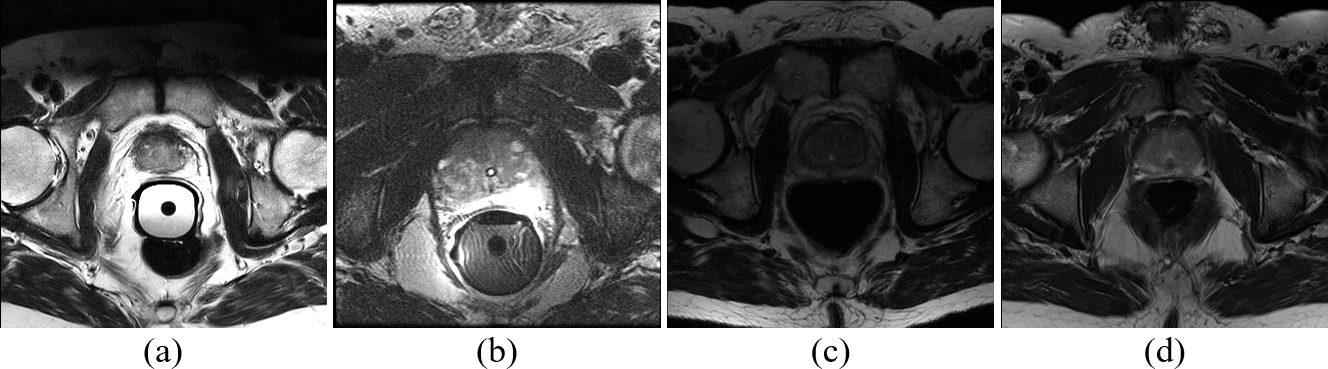}
	\caption{Four transversal slices from the PROMISE12 dataset.
	The T2-weighted MRI images were collected from different centers and with different protocols:
	(a) Haukeland University Hospital, Siemens, 1.5T, with endorectal coil (ERC);
	(b) Beth Israel Deaconess Medical Center University Hospital, GE, 3.0T, with ERC;
	(c) University College London, Siemens, 1.5/3.0T, without ERC;
	(d) Radbound University Nijmegen Medical Centre, Siemens, 3.0T, without ERC.
	 }
	\label{fig:data_promise12}
\end{figure}

\begin{figure}[t]
	\centering
	\includegraphics[width=0.99\linewidth]{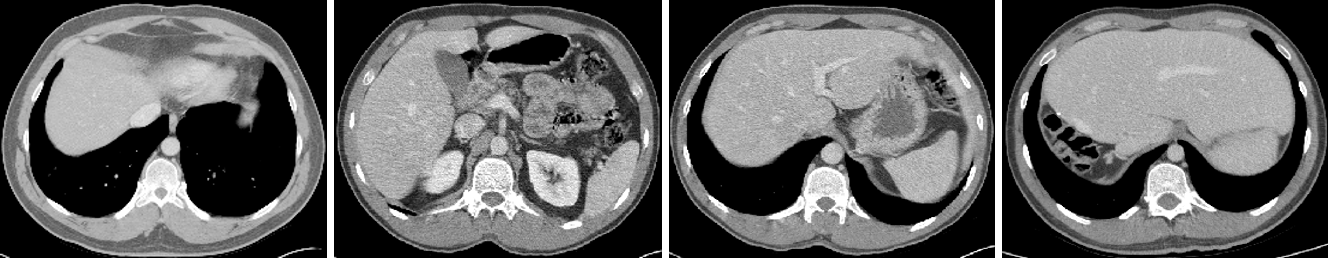}
	\caption{Example liver CT slices from the CHAOS dataset.
	The images show low tissue contrast and ambiguous boundaries of the liver regions.
	}
	\label{fig:data_chaos}
\end{figure}

\subsection{Attention Merge Module}
\label{ssec:amm}
The aforementioned embedding distance utilizes the context between neighboring slices as soft cue to track the variation of the volumetric shape, here we further selectively merge $P_{n'}$ and the original prediction $P_{n}$ to produce the final segmentation.

Inspired by SE Net~\citep{hu2018squeeze}, we propose an attention merge module (AMM), which makes the network pay more attention to the prediction that is more beneficial to the final segmentation.
As shown in Fig.~\ref{fig:merge}, the AMM mainly consists of three sequential functions of $F_{sq}$, $F_{score}$, $F_{mul}$.
Firstly, $F_{sq}$ squeezes global spatial information of the two predictions into a discrete feature vector, which employs the operation of global average pooling (GAP)~\citep{lin2013network}:
\begin{equation}
\label{formula:gap}
v = F_{sq}(u) = \frac{1}{W \times H}\sum_{i=1}^{W}\sum_{j=1}^{H}u(i,j),
\end{equation}
\textcolor{O}{where $u$ denotes the concatenation of  $P_n$ and $P_{n'}$}, $W$ and $H$ represent its dimension in width and height, respectively.
In order to fully capture the dependency between predictions, we then employ a simple gating mechanism with a sigmoid activation ($\sigma$):
\begin{equation}
\label{formula:4}
s = F_{score}(v, W_1, W_2) = \sigma(W_2\delta(W_1v)),
\end{equation}
where $\delta$ refers to ReLU, $W_1\in \mathbb{R}^{2r \times 2}$ and $W_2\in \mathbb{R}^{2 \times 2r}$.
To capture more abundant information, two fully-connected (FC) layers are used to enlarge the channel dimension by $r$ times.
Finally, the learned $s$ is used to re-weight the $P_{n'}$ and $P_{n}$:
\begin{equation}
\label{formula:5}
F_{mul}(u, s) = s\cdot u.
\end{equation}
where $u$ denotes the $P_{n'}$ and $P_{n}$.

Our AMM assigns dependency to $P_{n'}$ and $P_{n}$ to provide more comprehensive information for the following segmentation convolutions to generate the final prediction.

\section{Experiments and Results}
\label{sec:exper}
\subsection{Experimental Data and Pre-processing}
\label{ssec:data}
The proposed method was evaluated on two public medical image datasets:
the prostate MRI dataset (PROMISE12)~\cite{Litjens2013Evaluation}, and the liver CT dataset (CHAOS)\footnote{\url{https://chaos.grand-challenge.org/}}.
Both datasets are popular yet challenging benchmarks to evaluate the efficacy of segmentation algorithms.

The prostate MRI images from the PROMISE12 dataset were collected from four different centers, thus possessing diverse appearances (see Fig.~\ref{fig:data_promise12}).
We used 50 MRI volumes to conduct \textcolor{B}{patient-wise} five-fold cross validation.
The intra- and inter-slice resolutions of these volumes ranged from 0.25 mm to 0.75 mm, and from 2.2 mm to 4.0 mm, respectively. \textcolor{B}{Each 3D volume comprised between 15 and 54 slices.}
All MRI volumes were resampled to a fixed resolution of 1$\times$1$\times$1.5 mm$^{3}$.
The N4 bias field correction~\cite{tustison2010n4itk} was conducted, and then the intensity values were normalized to zero mean and unit variance.

The liver CT volumes were from the challenging CHAOS dataset.
Fig.~\ref{fig:data_chaos} illustrates some example CT images, which show low tissue contrast and ambiguous organs' boundaries.
We used 40 CT volumes to conduct \textcolor{B}{patient-wise} five-fold cross validation.
All CT slices were resampled to the size of 256$\times$256. \textcolor{B}{Each 3D volume comprised between 81 and 266 slices.}
The intensity values were clipped to an interval of [-100, 200] HU, and then normalized to zero mean and unit variance.
During training, contrast stretching, brightness stretching and elastic deformation were applied for data augmentation.

\begin{algorithm}
    \caption{\textcolor{G}{Contextual Embedding Learning for Volumetric Image Segmentation using 2D Networks}}
    \KwRequire{Image $S_n$, label $G_n$($n$ represents n-th slice of volumetric image)}
        \KwRequire{Backbone features $B_n$, backbone prediction $P_n$, slice-wisely neighboring matching prediction $P_{n'}$, refined prediction $P_{n-CE}$, distance maps $D_n$}
    \KwRequire{Backbone model $f(\cdot)$ with parameters $\theta_{b}$}
    \KwRequire{Embedding space transform module $e(\cdot)$ with parameters $\theta_{e}$}
    \KwRequire{Slice-wisely neighboring matching module $s(\cdot)$ with parameters $\theta_{s}$}
    \KwRequire{Attention Merge Module $m(\cdot)$ with parameters $\theta_{m}$}
    Initialize: $\theta_{b}$, $\theta_{e}$, $\theta_{s}$ and $\theta_{m}$
    
    \For{t \emph{in} \emph{[1, num\_epochs]}}
    {
        \For{i \emph{in} \emph{[1, num\_patients]}}{
        
            \textbackslash \textbackslash Compute backbone features and backbone predictions
            
            ($B_{0}$, ..., $B_n$), ($P_{0}$, ..., $P_n$) = $f(S_{0}, ..., S_n)$, 

            \textbackslash \textbackslash Compute embedding features

            $E_n$ = $e(B_n)$
            
            \textbackslash \textbackslash Compute distance maps via Eq 1 and Eq 2
            
            $D_n$ = $d(E_{n-l}, E_{n}, P_{n})$

            \textbackslash \textbackslash Compute slice-wisely neighboring matching prediction

            $P_{n'}$ = $s(P_{n-l}, D_n, B_n)$
        
            \textbackslash \textbackslash Compute refined prediction using Equ 3, Equ 4 and Equ 5

            $P_{n-CE}$ = $m(P_{n}, P_{n'})$
            
            Compute Loss $\mathcal{L}_{DSC}$

            Update $\theta_{b}$, $\theta_{e}$, $\theta_{s}$ and $\theta_{m}$ using SGD
            
        }
    }
    \Return $\theta_{b}$, $\theta_{e}$, $\theta_{s}$ and $\theta_{m}$
\end{algorithm}

\begin{table*}[t]
    \centering
    \small
    \caption{\textcolor{G}{Quantitative comparison of different methods on the prostate MRI dataset (Mean$\pm$SD, best results are highlighted in bold).
    ``*" indicates the results are statistically different with ours (Wilcoxon tests, $p$-value$<$0.05).}}
    \setlength{\tabcolsep}{2.0mm}{
        \begin{tabular}{l c c c c c c} 
            \toprule  
            Methods & DSC (\%) & ASSD (mm) & 95HD (mm)  & Sensitivity (\%) & Precision (\%)\\           
            \midrule
            V-Net~\citep{milletari2016v} & 85.12 $\pm$ 2.28*  & 1.95 $\pm$ 0.61* & 6.13 $\pm$ 4.71& 91.22 $\pm$ 4.62&83.41 $\pm$ 8.27* \\
            \midrule
            U-Net~\citep{ronneberger2015u} & 85.67 $\pm$ 6.89*  & 2.05 $\pm$ 2.36* & 6.92 $\pm$ 10.88 &85.94 $\pm$ 6.55* &85.05 $\pm$ 9.40*\\
            3D U-Net~\citep{cciccek20163d} & 86.03 $\pm$ 6.79 & 2.07 $\pm$ 2.08* & 6.70 $\pm$ 11.17&\textbf{89.26 $\pm$ 5.94}&\textbf{87.73 $\pm$ 7.04}\\    
            U-Net+CE (Ours) & \textbf{87.38 $\pm$ 4.42} & \textbf{1.55 $\pm$ 0.76} & \textbf{5.14 $\pm$ 4.01} &88.50 $\pm$ 4.07&87.73 $\pm$ 8.58\\
            \midrule
            DeepLabV3~\citep{Chen2017Rethinking} & 83.37 $\pm$ 6.21* & 2.13 $\pm$ 1.01* & 7.34 $\pm$ 6.48* &80.52 $\pm$ 7.63&\textbf{89.80 $\pm$ 9.19}\\
            3D DeepLabV3 & 84.86 $\pm$ 6.49 & \textbf{1.88 $\pm$ 0.82} & \textbf{4.83 $\pm$ 2.33} &\textbf{84.83 $\pm$ 6.93}&84.09 $\pm$ 8.83\\
            DeepLabV3+CE (Ours) & \textbf{84.94 $\pm$ 5.80} & 1.94 $\pm$ 0.91 & 6.26 $\pm$ 5.14&82.31 $\pm$ 5.41&87.27 $\pm$ 9.35\\
            \midrule
            Missformer~\citep{huang2022missformer}&82.81 $\pm$ 7.75*&2.13 $\pm$ 1.14&5.97 $\pm$ 3.25*&85.89 $\pm$ 6.55&82.22 $\pm$ 14.62\\
            SwinUNETR~\citep{hatamizadeh2021swin}&\textbf{87.74 $\pm$ 3.32}*&\textbf{1.31 $\pm$ 0.27}*&\textbf{4.11 $\pm$ 0.71}&\textbf{91.45 $\pm$ 3.22}*&\textbf{84.91 $\pm$ 7.60}\\
            Missformer+CE (Ours) &83.70 $\pm$ 6.39&2.19 $\pm$ 1.25&4.80 $\pm$ 3.79&86.00 $\pm$ 6.54&83.74 $\pm$ 13.47\\
            \bottomrule
    \end{tabular}}
    \label{table:result_promise12}
\end{table*}

\subsection{Implementation Details}
\label{ssec:implementation}
The segmentation model was implemented on the open source platform Pytorch.
The values of neighboring interval $l$, local patch size $k$, coefficient $r$ were empirically set to 1, 3, 64, respectively.
Dice loss was utilized to train the model:
\textcolor{B}{
\begin{equation}
\label{formula:6}
\mathcal{L}_{\mathcal{DSC}}=1-\frac{1}{C}\sum_{c=1}^{C}\frac{2\sum_{i=1}^{N}{y_{i,c}p_{i,c}}}{\sum_{i=1}^{N}{y_{i,c}+\sum_{i=1}^{N}p_{i,c}}},
\end{equation}
where $y_{i,c}$ employed a one-hot encoding scheme representing the ground truth labels, $p_{i,c}$ denoted the predicted values produced by the model for each class, and where indices i and c iterated over all pixels and classes, respectively.}
Stochastic gradient descent (SGD) was used with batch size of 50, learning rate of 0.01, momentum of 0.9 and weight decay of 0.001 to control the learning progress of the network.
The total number of training epochs was 300.
All experiments were conducted on a single NVIDIA GeForce GTX 2080 GPU.

\textcolor{G}{The Algorithm 1 provides the pseudo-code of our proposed method.}

\subsection{Evaluation Metrics}
\label{ssec:metrics}
The metrics employed to quantitatively evaluate the segmentation accuracy included Dice similarity coefficient (DSC), average symmetric surface distance (ASSD),
95\% Hausdorff distance (95HD),
\textcolor{G}{sensitivity and precision}~\citep{Taha2015Metrics}.
All above metrics were calculated in 3D space.
DSC is the primary metric used for assessing the segmentation accuracy,
which measures the relative volumetric overlap between the predicted and ground-truth
segmentations.
ASSD determines the average distance between the surfaces of the predicted and ground-truth segmentations.
95HD is similar to ASSD but more sensitive to the localized disagreement as it determines the 95th percentile of all calculated Hausdorff distances.
\textcolor{G}{
Sensitivity indicates the model's capability to detect the foreground class in an image.
Precision indicates the model's capability to predict correct positive samples in an image.}
A better segmentation shall have smaller ASSD and 95HD, and larger values of DSC, precision and sensitivity.

\subsection{Segmentation Accuracy}
\label{ssec:exp}
To demonstrate the efficacy of the proposed CE block, we applied it to enhance three popular 2D segmentation networks, i.e., U-Net~\citep{ronneberger2015u}, DeepLabV3~\citep{Chen2017Rethinking} and \textcolor{G}{Missformer~\citep{huang2022missformer}}, in a plug-and-play manner.
We compared the CE-enhanced 2D networks with their 3D version, i.e., 3D U-Net~\citep{cciccek20163d}, 3D DeepLabV3 and \textcolor{G}{SwinUNETR~\citep{hatamizadeh2021swin}}.
We also compared our method with two well-established models for prostate MRI segmentation (V-Net~\citep{milletari2016v}) and liver CT segmentation (U-Net++~\citep{zhou2018unet++}), respectively.

\begin{figure*}[t]
    \centering
    \includegraphics[width=0.78\linewidth]{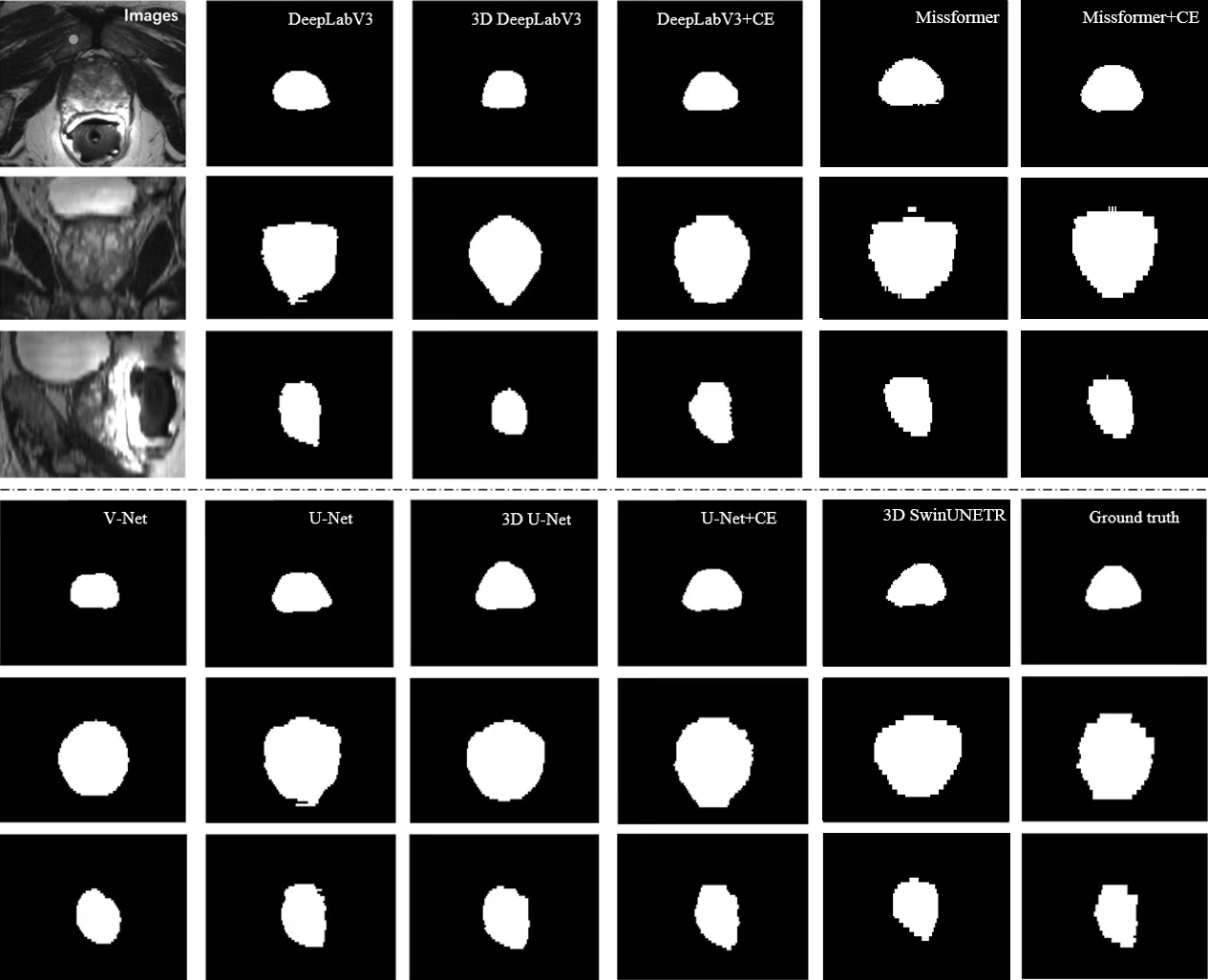}
    \caption{Qualitative illustration of prostate MRI segmentation results from different methods in transversal, coronal, and sagittal views, respectively.
    The proposed CE block boosts the conventional 2D networks to generate more continuous and accurate results in 3D space.}
    \label{fig:promise_result}
\end{figure*}

Table~\ref{table:result_promise12} reports the numerical results of all methods on the prostate MRI dataset.
To investigate the statistical significance of the proposed method over compared methods on each metrics, the Wilcoxon signed-rank test was employed and also reported in Table~\ref{table:result_promise12}.
It can be observed that our CE-enhanced 2D networks can be regarded as significantly better than the conventional 2D networks on the DSC evaluation metric.
Furthermore, regarding the backbones of U-Net and DeepLabV3,
the $p$-values of CE-enhanced 2D networks $vs$. 3D networks on almost all metrics are beyond the 0.05 level, which indicates that the CE-enhanced 2D networks achieved similar performance as the 3D U-Net and 3D DeepLabV3 did.
However, due to the superior performance of the cutting-edge 3D Transformer,
the Missformer+CE could not attain similar segmentation results compared to the 3D SwinUNETR.
Fig.~\ref{fig:promise_result} visualizes the segmentation results in transversal, coronal and sagittal views, respectively.
Apparently, our CE-enhanced networks obtained similar segmented boundaries to the ground truth.
Compared with conventional 2D networks, our method generated more accurate and continuous results in volumetric view (see the coronal and sagittal views in Fig.~\ref{fig:promise_result}).
In general, the results shown in Table~\ref{table:result_promise12} and Fig.~\ref{fig:promise_result} prove the effectiveness of our contextual embedding strategy.
Compared with conventional 2D networks, the proposed CE block leveraged the learned embedding and the slice-wisely neighboring matching as an internal guidance to transfer contextual information thus facilitating 2D networks to capture volumetric representation properly.

\begin{table*}[t]
    \centering
    \small
    \caption{\textcolor{G}{Quantitative comparison of different methods on the liver CT dataset (Mean$\pm$SD, best results are highlighted in bold).
    ``*" indicates the results are statistically different with ours (Wilcoxon tests, $p$-value$<$0.05).}}
    \setlength{\tabcolsep}{2.0mm}{
        \begin{tabular}{l c c c c c c} 
            \toprule  
            Methods & DSC (\%) & ASSD (mm) & 95HD (mm) & Sensitivity (\%) & Precision (\%)\\           
            \midrule
            UNet++~\citep{zhou2018unet++} & 94.23 $\pm$ 2.94* & 0.95 $\pm$ 0.58*  & 4.89 $\pm$ 6.50* & 96.05 $\pm$ 1.96 & 96.43 $\pm$ 1.39\\
            \midrule
            U-Net~\citep{ronneberger2015u} & 94.13 $\pm$ 2.32* & 1.84 $\pm$ 1.12*  & 8.62 $\pm$ 7.62* & 96.14 $\pm$ 1.47 & 96.54 $\pm$ 1.08\\
            3D U-Net~\citep{cciccek20163d} & 95.67 $\pm$ 2.23* & 2.31 $\pm$ 1.72* & 2.31 $\pm$ 1.72 & \textbf{97.14 $\pm$ 0.94} & 96.38 $\pm$ 0.86\\    
            U-Net+CE (Ours) & \textbf{96.49 $\pm$ 1.01} & \textbf{0.50 $\pm$ 0.17} & \textbf{1.77 $\pm$ 0.84} & 96.21 $\pm$ 1.52 & \textbf{96.75 $\pm$ 1.07}\\
            \midrule
            DeepLabV3~\citep{Chen2017Rethinking} & 95.05 $\pm$ 4.85* & 1.78 $\pm$ 2.24*  & 7.62 $\pm$ 14.83* & 94.98 $\pm$ 1.30 & 95.84 $\pm$ 0.93\\
            3D DeepLabV3 & 95.33 $\pm$ 1.44 & 1.49 $\pm$ 0.81* & 5.83 $\pm$ 5.56* 
            & 94.91 $\pm$ 1.55 & 94.55 $\pm$ 1.29*\\
            DeepLabV3+CE (Ours) & \textbf{95.59 $\pm$ 2.03} & \textbf{0.61 $\pm$ 0.37} & \textbf{1.95 $\pm$ 1.50} & \textbf{95.88 $\pm$ 1.00} & \textbf{96.09 $\pm$ 0.74}\\
            \midrule
            Missformer~\citep{huang2022missformer}&96.45 $\pm$ 0.40&1.14 $\pm$ 0.20&
            3.81 $\pm$ 0.81&96.58 $\pm$ 1.39&96.36 $\pm$ 1.06\\
            SwinUNETR~\citep{hatamizadeh2021swin}&\textbf{96.63 $\pm$ 0.46}&\textbf{1.10 $\pm$ 0.20}&\textbf{3.24 $\pm$ 0.50}&	\textbf{97.14 $\pm$ 1.43}&
            96.17 $\pm$ 1.27\\
            Missformer+CE (Ours) &96.56 $\pm$ 0.42&1.11 $\pm$ 0.15&3.55 $\pm$ 0.63&	
            96.53 $\pm$ 1.47&\textbf{96.62 $\pm$ 0.96}\\
            \bottomrule
    \end{tabular}}
    \label{table:result_chaos}
\end{table*}

\begin{table*}
    \centering
    \small
    \caption{\textcolor{O}{Efficiency comparison between baseline 2D networks, corresponding 3D versions, and the CE-enhanced 2D networks in terms of parameter amount and floating point operations (FLOPs).}}
    \setlength{\tabcolsep}{5.5mm}{
        \begin{tabular}{l c c c} 
            \toprule  
             Methods & $\#$ Params & FLOPs (G) & Input size\\         
            \midrule
            UNet~\citep{ronneberger2015u} & 7.76 M & 3.42 & (1, 1, 128, 128)\\
            3D UNet~\citep{cciccek20163d} & 5.64 M & 61.97 & (1, 1, 64, 128, 128)\\
            U-Net+CE (Ours) & 7.96 M & 6.62 & (1, 1, 128, 128)\\
            DeepLabV3~\citep{Chen2017Rethinking} & 25.42 M & 6.68 & (1, 1, 128, 128)\\ 
            3D DeepLabV3 & 74.72 M & 322.17 & (1, 1, 64, 128, 128)\\
            DeepLabV3+CE (Ours) & 25.63 M & 7.66 & (1, 1, 128, 128)\\
            Missformer~\citep{huang2022missformer} & 10.77 M & 6.95 & (1, 1, 224, 224)\\    
            3D Swin unetr~\citep{hatamizadeh2021swin} & 15.7 M & 12.71 & (1, 1, 64, 64, 64)\\
            Missformer+CE (Ours) & 10.97 M & 27.36 & (1, 1, 224, 224)\\
            \bottomrule
        \end{tabular}}
    \label{table:efficiency}
\end{table*}

\begin{figure*}[t]
    \centering
    \includegraphics[width=0.75\linewidth]{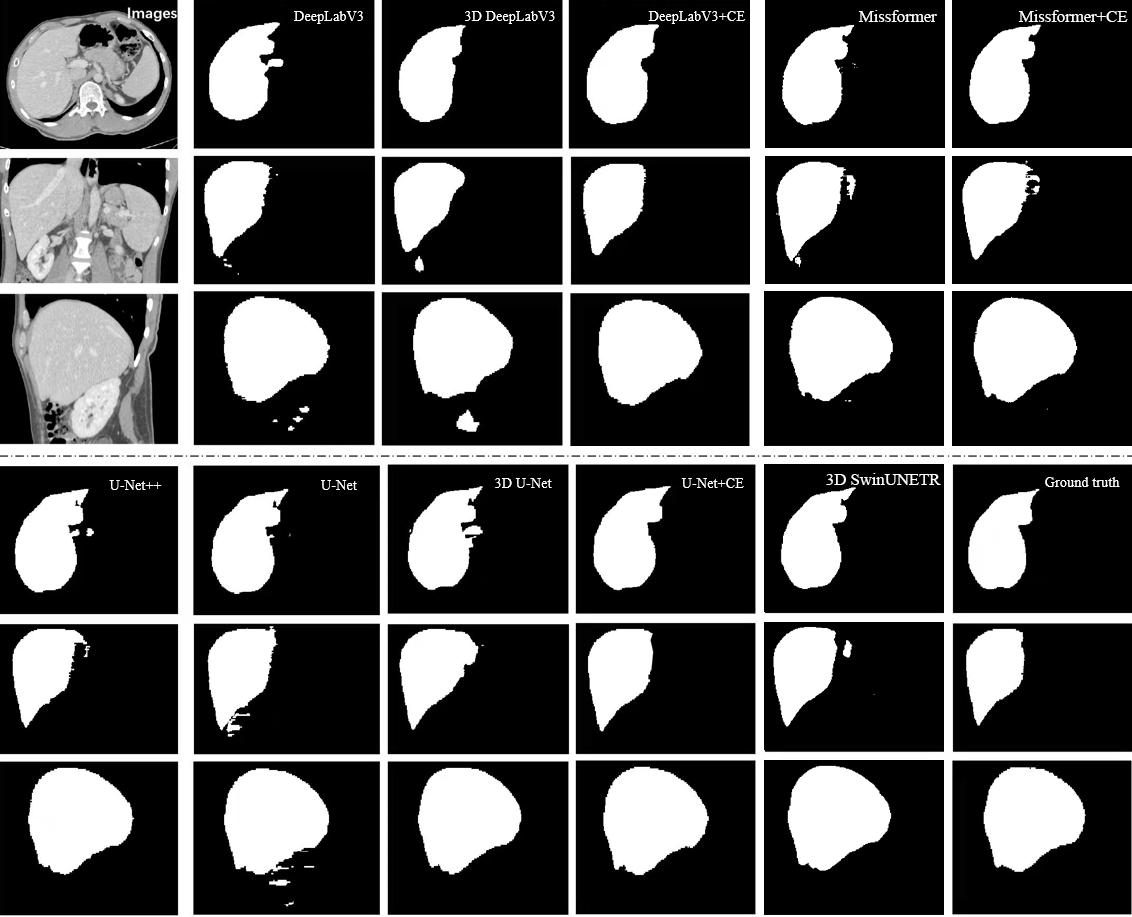}
    \caption{\textcolor{B}{Qualitative illustration of liver CT segmentation results from different methods in transversal, coronal, and sagittal views, respectively.
    Our CE-enhanced networks generate the most similar segmentation to the ground truth.}}
    \label{fig:chaos_result}
\end{figure*}

The quantitative results on the liver CT dataset are listed in Table~\ref{table:result_chaos}.
It can be observed that the CE-enhanced networks outperformed conventional 2D networks on all metrics.
It is worth noting that the proposed CE block largely improved the 2D U-Net and DeepLabV3 with respect to the metrics of ASSD and 95HD, corroborating that the CE block has strong capability to utilize the slice-wise context for the accurate volumetric segmentation.
The statistical results in Table~\ref{table:result_chaos} further shows our method can make the segmentation performance of the 2D networks close to or even surpass that of the 3D networks.
Figs.~\ref{fig:chaos_result} visualizes the liver segmentation results in transversal, coronal and sagittal views, respectively.
Our method can successfully infer the ambiguous boundaries and attain an accurate and smooth volumetric segmentation.

\subsection{Efficiency Comparison}
\label{ssec:efficiency}
We further compared the baseline 2D networks, corresponding 3D versions, and the CE-enhanced 2D networks in terms of parameter amount and floating point operations (FLOPs).
As shown in Table~\ref{table:efficiency}, it can be observed that compared to the 2D backbone networks, the parameter number of the proposed CE Block is relatively small (about 0.2 M).
In addition, the CE-enhanced U-Net and DeepLabV3 are more lightweight compared with the corresponding 3D networks in terms of FLOPs.
The comparison results in Table~\ref{table:efficiency} demonstrate the proposed CE block is a memory-efficient solution to enhance the 2D networks for volumetric segmentation.

\section{Discussion}
\label{sec:discussion}
The automatic segmentation of organs in volumetric medical images, such as MRI and CT scans, plays an important role in computer-assisted diagnosis and interventions.
\textcolor{G}{For example, it facilitates the calculation of the required radiotherapy dose for specific region-of-interests. In addition, the segmentation results can be further used to assist mask/surface-based medical image registration.}
Conventional 2D convolutional networks treat volumetric images as a stack of 2D slices thus cannot learn spatial information in the third dimension.
The 3D CNNs are beneficial to extract richer volumetric information, but often require high computation cost and severe GPU memory consumption.

\begin{figure}[t]
	\centering
	\includegraphics[width=\linewidth]{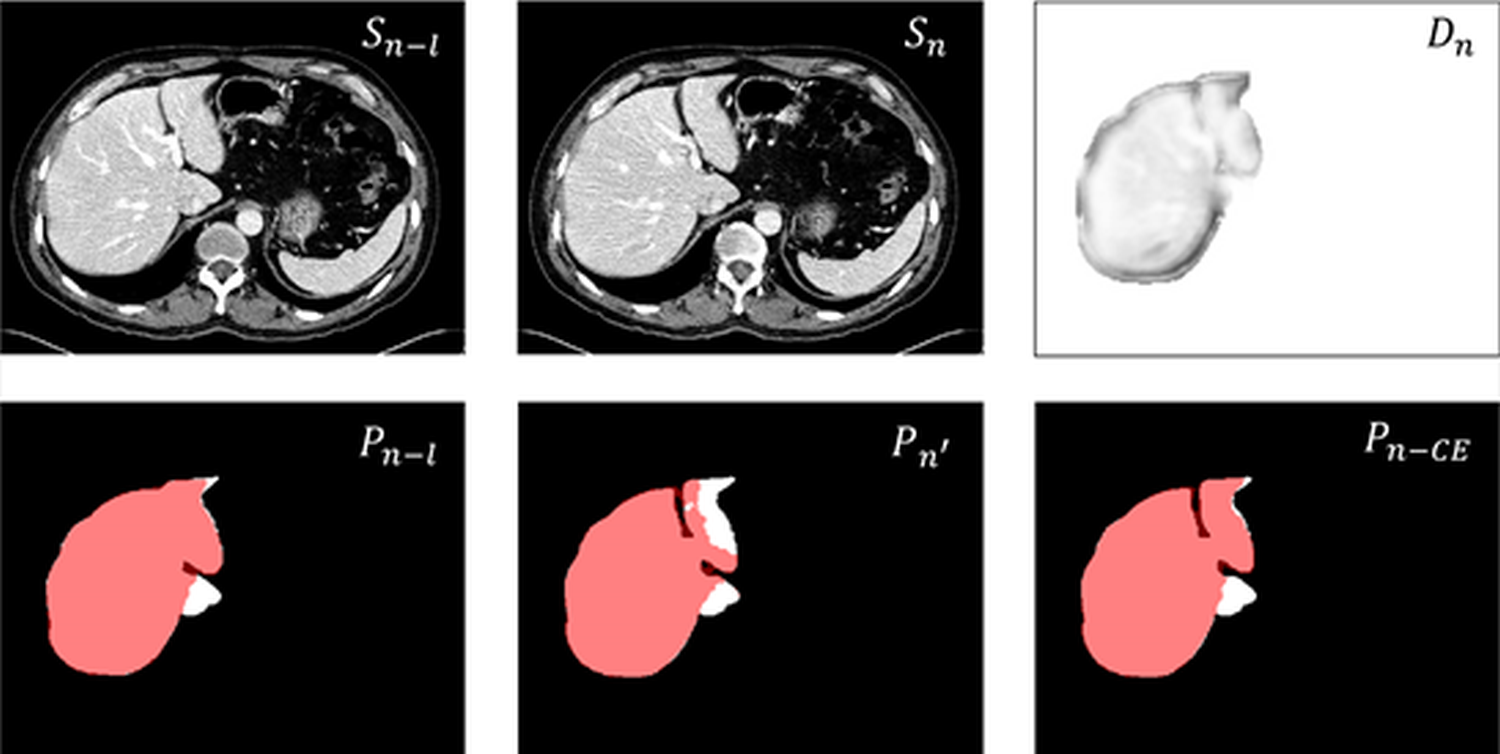}
	\caption{\textcolor{G}{An example to visualize the distance map in the embedding space enhancing the segmentation.}}
	\label{fig:discussion}
\end{figure}

To alleviate above issues, the proposed contextual embedding strategy can convert conventional 2D segmentation network consisting of the encoder-decoder architecture into a pseudo-3D segmentation network that captures the slice-wise context, which solves the issue in a memory-friendly way that the segmentation results of 2D networks are discontinuous and incomplete in the 3D space.
\textcolor{G}{
Our assumption is that in the embedding space, pixels of the same type exhibit similarity, reflected in the similarity of distances calculated by Eq~(\ref{formula:distanceP2P}). The resulting distance map positively impacts the segmentation results by providing valuable feedback to help the network distinguish between pixels of the same type. Conversely, if the network incorrectly assigns pixels of different types to the same category, it adversely affects the results. These feedbacks contribute to optimizing the network. In summary, dividing pixels into the same category or different categories is a process of network adaptation.
As shown in Fig.~\ref{fig:discussion}, segmentation of image $S_n$ without using CE Block yields the result $P_{n'}$, where the red region is the prediction and the white mask denotes ground truth. In contrast, $D_n$ is the distance map between $S_n$ and $S_{n-l}$, and $P_{n-l}$ is the refined prediction of image $S_{n-l}$. It can be observed that the upper right part of the liver in the distance map suggests similarity to the previous frame in the embedding space, indicating that this region should be segmented as part of the liver. We refine $P_{n'}$ using $D_n$ to obtain better result $P_{n-CE}$ for $S_n$.
}

It is worth noting that the proposed contextual embedding mechanism is employed as an internal guidance of the conventional networks, but not as a direct constraint for the final segmentation prediction.
Thus the whole network can be trained without requiring an extra supervision on the embedding.
In addition, our CE block is a plug-and-play module without too much additional convolutions.
Compared with 3D convolutional networks, our CE-enhanced 2D networks are more computation-/memory-efficient for volumetric segmentation, as shown in Table~\ref{table:efficiency}.

In this study, we aim to design a novel module to enhance the 2D convolutional networks with contextual information for more accurate and efficient volumetric segmentation.
Thus we mainly pay attention to design a lightweight and easy-to-use module rather than breaking through the state-of-the-art segmentation accuracy.
The proposed module is expected to be integrated into the current 2D encoder-decoder architectures for their performance improvements.
Therefore, in our experiments, we focused on comparing our method with the popular 2D backbones rather than with the current cutting-edge networks on the two experimental datasets.

\textcolor{P}{While our method demonstrates improved results with the original 2D network, it does not optimize the backbone network itself. Therefore, the segmentation performance may still subject to the backbone architecture. As shown in Fig.~\ref{fig:discussion}, although the proposed CE block can refine the segmentation results, there still exists under-segmentation region due to the inherent capability of the backbone network. Additionally, in this study, our method has only been validated on two single-organ segmentation tasks. Its performance across a wider range of anatomical structures, or its adaptability to different imaging modalities (i.e., ultrasound images) could be interesting future directions.}

\section{Conclusion}
\label{sec:con}
For the volumetric segmentation, conventional 2D networks may hardly exploit spatial information in 3D, while most 3D networks suffer from occupying excessive computational resources.
In this study, we present a novel plug-and-play contextual embedding block to enhance the 2D networks for volumetric segmentation.
Specifically, the proposed CE block leverages the learned embedding and the slice-wisely neighboring matching to aggregate contextual information.
The CE block has neither excessive convolutions nor 3D complicated algorithms, which is an efficient solution to boost the volumetric representation of the 2D networks.
Experimental results on challenging prostate MRI dataset and abdominal CT dataset show that the proposed method effectively leverages the inter-slice context and consistently improves 2D networks' performance on volumetric images.

\section*{Acknowledgements}
This work was supported in part by the National Natural Science Foundation of China under Grant 62071305,
in part by the Guangdong-Hong Kong Joint Funding for Technology and Innovation under Grant 2023A0505010021,
in part by the Guangdong Basic and Applied Basic Research Foundation under Grant 2022A1515011241,
and in part by the Shenzhen Municipal Scheme for Basic Research under Grant JCYJ20210324100208022.




  \bibliographystyle{model2-names} \biboptions{authoryear}
  \bibliography{ref}


%
%
%

\end{document}